\documentclass[twocolumn,prd,aps,showpacs,showkeys,amsmath,amssymb,nofootinbib]{revtex4-1}
\usepackage{bm}

\usepackage{color}
\usepackage{amsmath}
\usepackage{amsfonts}
\usepackage{verbatim}
\usepackage{amssymb}
\usepackage{graphicx}
\usepackage{epstopdf}
\usepackage{mathrsfs}
\usepackage{epsfig}
\usepackage{slashed}
\usepackage{bbold}
\usepackage{color} 

\providecommand{\U}[1]{\protect\rule{.1in}{.1in}}
\newcommand{\bl}{\boldsymbol}

\newcommand{\ph}{\phantom}

\begin{document}


\title{A Class of Integrable Metrics II }
\author{Gabriel Luz Almeida and Carlos Batista}
\email[]{carlosbatistas@df.ufpe.br}
\affiliation{Departamento de F\'{\i}sica, Universidade Federal de Pernambuco,
Recife, Pernambuco  50740-560, Brazil}


\begin{abstract}
Starting with a subclass of the four-dimensional spaces possessing two commuting Killing vectors and a non-trivial Killing tensor, we fully integrate Einstein's vacuum equation with a cosmological constant. Although most of the solutions happen to be already known, we have found a solution that, as far as we could search for, has not been attained before. We also characterize the geometric properties of this new solution, it is a Kundt spacetime of Petrov type II possessing a null Killing vector field and an isometry algebra that is three-dimensional and abelian. In particular, such solution becomes a $pp$-wave spacetime when the cosmological constant is set to zero.
\end{abstract}
\keywords{Exact solutions; Killing tensor; General relativity; Bianchi type I; Kasner solution}
\maketitle

\section{Introduction}

Due to the nonlinearity of Einstein's equation, it is virtually impossible to integrate it analytically without imposing restrictions over the initial ansatz. The most common way of doing so is by the imposition of symmetries. For instance, Schwarzschild solution has been found assuming that the spacetime has spherical symmetry, namely it has three Killing vectors whose  Lie algebra is $\mathfrak{so}(3)$. Likewise, Kerr solution has been obtained relying on the existence of two commuting Killing vectors \cite{Kerr}, i.e. the spacetime was assumed to be stationary and axisymmetric. It is important to keep in mind that the hypothesis of two commuting Killing vectors is not over-restrictive from the physical point of view, since the rigidity theorem states that the equilibrium state of an astronomical object should be stationary and axisymmetric \cite{HawkingRigidity,Chrusciel:1996bj}.

Besides the symmetries of the spacetime, which are generated by Killing vectors, one can also impose symmetries on the geodesic motion, which are generated by Killing tensors and Killing-Yano tensors \cite{Carter-KleinG,Santillan}.  Since the metric is always a Killing tensor, the existence of an extra Killing tensor along with two independent Killing vectors leads to four first integrals for the geodesic motion, which enables full integrability. Nevertheless, one might wonder whether it is plausible to assume the existence of a Killing tensor in physical spacetimes. The known examples tell us that the answer is yes. For instance, four-dimensional Kerr metric and, more generally, Kerr-NUT-(A)dS spacetimes in arbitrary dimension \cite{KerrNutAds}, are all endowed with enough Killing tensors to allow the integrability of the geodesic motion \cite{Kubiz,Krtous}. Thus, some of the most physically important exact solutions for Einstein's vacuum equation are endowed with Killing tensors. In addition to being related to the integrability of the geodesic motion, these Killing tensors are also related to the integrability of field equations in such spacetimes, like scalar fields \cite{Frol-KG}, electromagnetic fields \cite{KrtousMaxwell,Teukolsky}, and spin 1/2 fields \cite{OotaDirac}.  Probably, the existence of these objects might also be related to the integrability of Einstein's equation itself \cite{Yasui}, as hinted by the successful integration of gravitational perturbations through the use of Killing tensors \cite{OotaGrav}. Moreover, these Killing and Killing-Yano tensors can play an important role in supersymmetric theories \cite{KY-SUSY,Cariglia}.

With these motivations in mind, in the present article we will search for solutions of Einstein's vacuum equation with a cosmological constant within the class of spacetimes possessing a Killing tensor and two commuting Killing vectors. The general form of the spaces with such symmetry properties has been found by Benenti and Francaviglia in  Ref. \cite{BenentiFrancaviglia} and is given by:
\begin{align}
 g^{ab}\partial_{a}\partial_{b}  =\frac{1}{S_{x}+S_{y}}\,& \Big[
\,G_{x}^{ij}\,\partial_{\sigma_i}\partial_{\sigma_j}\,+\,G_{y}^{ij}\,\partial_{\sigma_i}\partial_{\sigma_j}    \nonumber\\
& \;\;\;\quad  +   \Delta_{x}\,   \partial_{x}^{2}+\Delta_{y}\, \partial_{y}^{2}\, \Big]  \, , \label{BFmetric}
\end{align}
where functions with subscript $x$ are arbitrary functions of $x$, while those with subscript $y$ are arbitrary functions of $y$. For instance, $\Delta_{x} = \Delta_{x}(x)$. The indices $i,j$ run through $\{1,2\}$ and label the cyclic coordinates $\sigma_1$ and $\sigma_2$. Note that we can assume that  $G_{x}^{ij} = G_{x}^{ji}$ and $G_{y}^{ij} = G_{y}^{ji}$, due to the symmetry of the metric. The rank two Killing tensor associated to this metric is given by
\begin{align}
  \boldsymbol{K}\,=\,\frac{1}{S_{x}+S_{y}}\,\Big[& \,S_{x}\,G_{y}^{ij}\, \partial_{\sigma_i}\partial_{\sigma_j}
+ S_{x}\,\Delta_{y} \,\partial_{y}^{2}    \nonumber\\
&  - S_{y}\,G_{x}^{ij}\,\partial_{\sigma_i}\partial_{\sigma_j}  - S_{y} \,\Delta_{x} \,  \partial_{x}^{2}  \, \Big]  \,. \label{KillingT1}
\end{align}

In recent previous works we have already exploited the integrability of Einstein's equation of some spaces within the class of metrics (\ref{BFmetric}). In Ref. \cite{AnabalonBatista}, one of us (C.B.) along with A. Anabal\'{o}n investigated the subcase in which the determinants of the matrices $G_x^{ij}$ and $G_y^{ij}$ are both zero. It has been found that  Einstein's vacuum equation with a cosmological constant is fully integrable for such a subcase, with Kerr-NUT-(a)dS being a particular solution. Latter, the present authors also considered the subcase of vanishing determinant for $G_x^{ij}$ and $G_y^{ij}$ but, instead of vacuum,
a gauge field of arbitrary gauge algebra have been considered as a source for the gravitational field \cite{GabrielBatista}. In particular, new exact solutions have been attained in Ref. \cite{GabrielBatista}.

Now, the idea is to explore another subcase of the class of spaces (\ref{BFmetric}). Namely, the one in which one of the matrices $G_{x}^{ij}$ or $G_{y}^{ij}$ vanishes identically. For definiteness, we shall assume $G_x^{ij} = 0$. In this case it is immediate to notice that the line element is given by
\begin{equation}\label{metric1}
  ds^2 = (S_x + S_y) \left[ H_y^{ij}\,d\sigma_i d\sigma_j + \frac{dx^2}{\Delta_x} + \frac{dy^2}{\Delta_y}  \right] \,,
\end{equation}
where $H_y^{ij}$ are arbitrary functions of $y$. Note that in the general case, when  $G_{x}^{ij}$ and $G_{y}^{ij}$ are both nonzero, the line element would have the same algebraic structure above, but the components $H^{ij}$ would be convoluted combinations of functions of $x$ and functions of $y$. As we shall see in the sequel, Einstein's vacuum equation for the class of spaces described by (\ref{metric1}) is integrable. It will be shown that although most of the solutions found within this class are already known, we arrive at a particular solution that, as far as the authors know, has not been attained before.

The outline of the article is the following. At the next section we start the integration of Einstein's equation and conclude that the calculations should be split in three different cases depending on the constancy of the functions $S_x$ and $S_y$. The case in which both functions are constant is tackled in subsection  \ref{SubSecA}, which yields flat spaces as the only solutions. Then, the case in which $S_y$ is constant while $S_x$ is non-constant is considered in subsection \ref{SubSecB}, with the only solutions being spaces of constant curvature. Finally, the case in which just $S_x$ is constant is considered in subsection \ref{SubSecC}. During the integration process of the latter case we conclude that there is a special subcase that must be considered separately. In \ref{SubSecC1} we treat the general case and arrive at a generalization of Kasner spacetime, while the special subcase is tackled in subsection \ref{SubSecC2}  and leads to a solution that, as far as the authors know, has not been described in the literature yet. Then, in Sec. \ref{Sec.NewSOL} we investigate the geometrical features of the new solution. We show that this solution is a Kundt spacetime of Petrov type II possessing a null Killing vector field  and that it reduces to a $pp$-wave spacetime when the cosmological constant vanishes. Its isometry algebra is three-dimensional and abelian, so that it is Bianchi type I, but, differently from the most known solutions of this type, the line element cannot be diagonalized using the cyclic coordinates associated to the Killing vectors. The regularity of the new solution and its asymptotic form are also investigated. The conclusions and perspectives are presented at Sec. \ref{Sec.Conc}. At appendix \ref{AppendixA} we show that at the asymptotic limit, the new solution goes to a Kasner spacetime.



\section{Integrating Einstein's Equation}\label{Sec.Integration}

The goal of this work is to integrate Einstein's field equation in vacuum with a cosmological constant $\Lambda$. That is, we want to find the most general solution of the equation
\begin{equation}\label{Einsteinseq}
R_{ab}=\Lambda g_{ab},
\end{equation}
for line elements of the form (\ref{metric1}), where $R_{ab}$ stands for the Ricci tensor.
Nevertheless, before doing so, it is useful to replace the three arbitrary functions $H_y^{11}$, $H_y^{22}$ and $H_y^{12}=H_y^{21}$ appearing in (\ref{metric1}) by the three functions $P_y$, $Q_y$ and $\Omega_y$ defined in a way that the line element assumes the following form:
\begin{align}\label{metric2}
ds^2=& \,S \,\Big( -\frac{1}{\Omega_y}d\sigma_1^2+ \frac{Q_y^2-P_y^2}{\Omega_y}d\sigma_2^2 \nonumber\\
& \quad \;\;+  \frac{2P_y}{\Omega_y} d\sigma_1 d\sigma_2  +\frac{dx^2}{\Delta_x}+\frac{dy^2}{\Delta_y} \,\Big),
\end{align}
where $S=S_x+S_y$. This represents no loss of generality.


Now, an immediate integration of the component $R_{\sigma_1}^{\ph{\sigma_1}\sigma_2}=0$  of Einstein's equation for the function $\Delta_y$ provides
\begin{equation}\label{Dy}
\Delta_y=\frac{c_1 \,Q_y^2 \,\Omega_y^2}{(S_x + S_y)^2(P_y')^2} \,,
\end{equation}
where $c_1$ is an arbitrary integration constant and the prime denotes a derivative with respect to the variable on which a function depends. Although Eq. (\ref{Dy}) is correct when $S_x$ is a constant function, such equation cannot be used when $S_x$ is non-constant, otherwise $\Delta_y$ would also depend on $x$. Thus, the case in which $S_x$ is a non-constant function of $x$ must be handled with special care\footnote{The equation $R_{\sigma_1}{}^{\sigma_2}=0$ has the structure $A_y + B_y S_x=0$, where $A_y$ and $B_y$ are functions of $y$.  If $S_x$ is constant the general solution is $A_y = -B_y S_x$. However, if $S_x$ is non-constant the general solution is $A_y=0$ and $B_y=0$, thus yielding an extra constraint.}. Doing so, we find that the equation $R_{\sigma_1}^{\ph{\sigma_1}\sigma_2}=0$  yields the following constraints:
\begin{equation}\label{Dy2}
\Delta_y=\frac{c_1 \,Q_y^2 \,\Omega_y^2}{(P_y')^2} \;\; \textrm{ and } \;\; S'_y =  0  \,. \;\;\; (\textrm{when } S'_x\neq 0)
\end{equation}
In order to attain both of the expressions \eqref{Dy} and \eqref{Dy2}, we have considered that $P_y'\neq 0$. The special case in which $P_y$ is constant will be considered latter.

Now, assuming either \eqref{Dy} or  \eqref{Dy2} to hold, and then integrating $R_{\sigma_2}^{\ph{\sigma_2}\sigma_1}=0$, we find that in both cases $Q_y$ must be given by
\begin{equation}\label{Qy}
Q_y=\sqrt{(P_y-a_1)(P_y-a_2)} \,,
\end{equation}
with $a_1$ and $a_2$ being arbitrary integration constants.

Also, irrespective of assuming the latter expressions for $\Delta_y$ and $Q_y$, the integration of the component $R_x{}^y=0$ leads to the following constraint:
\begin{equation*}
S_x'\,S_y'=0 \,.
\end{equation*}
Thus, we face three possible cases to be followed depending on whether the functions $S_x(x)$ and $S_y(y)$ are constant or not. Namely, (A) the functions $S_x$ and $S_y$ are both constant, (B) $S_x$ is non-constant and $S_y$ constant, and (C) $S_x$ is constant and $S_y$ non-constant. Particularly, note that in cases (A) and (C), $\Delta_y$ is given by Eq. \eqref{Dy}, while in the case (B) we must use Eq. \eqref{Dy2}. In the following section, each of these three cases will be treated separately. As we shall see, the cases (A) and (B) do not provide any particularly interesting solutions, while the case (C) will lead to solutions with richer physics: a generalization of Kasner solution, that is already available in the literature,  and a new solution of Petrov type II possessing a null Killing vector field and whose isometry algebra is three-dimensional and abelian.


\subsection{Subcase $S_x'=0$ and $S_y'=0$}\label{SubSecA}

In this subsection we investigate the simplest of the three possible cases regarding the constancy of the functions $S_x$ and $S_y$, namely we shall consider that they are both constant. This gives rise to a constant conformal factor $S$ which can be easily incorporated into the coordinates by a scaling transformation, so that we can set
\begin{equation*}
  S= S_x + S_y = 1 \,.
\end{equation*}

Then, assuming $S=1$, along with Eqs. (\ref{Dy}) and (\ref{Qy}) for $\Delta_y$ and $Q_y$, it follows that $R_{x}^{\ph{x}x}$ is automatically zero, so that the equation $R_{x}^{\ph{x}x} = \Lambda \delta^x_x$ states that the cosmological constant must vanish, $\Lambda = 0$. Then, integrating $R_{\sigma_1}^{\ph{\sigma_1}\sigma_1}=\Lambda=0$, we find
\begin{equation}\label{Omegay1}
\Omega_y=c_2\,(P_y-a_1)^{d} \, (P_y-a_2)^{1-d} \,,
\end{equation}
where $c_2$ and $d$ are arbitrary integration constants. In order to attain (\ref{Omegay1}) it was necessary to assume $a_1 \neq a_2$. Indeed, the special case $a_1=a_2$ would lead to a different expression for $\Omega_y$, but let us put this particular case aside and deal with it at the end of this subsection. With the latter expressions for $S$, $\Delta_y$, $Q_y$, and $\Omega_y$ at hand, the equation $R_{y}^{\ph{y}y}=\Lambda=0$ leads to the constraint $d = 0$.  Actually, another possibility for solving $R_{y}^{\ph{y}y}=0$ is $d=1$, but this is equivalent to $d=0$ when we interchange the arbitrary constants $a_1$ and $a_2$, so that we just need to consider $d=0$. Thus, $\Omega_y$ should be given by:
\begin{equation*}
 \Omega_y=c_2\,  (P_y-a_2) \,.
\end{equation*}
With this expression for $\Omega_y$ along with the latter expressions for $S$, $Q_y$ and $\Delta_y$, it follows that the Riemann tensor is identically zero. Thus, the solution is the flat space.  In particular, the Ricci tensor vanishes, so that we must have $\Lambda= 0$.

In the latter integration, we have excluded two possibilities, namely the case $a_1=a_2$ and the case in which  $P_y$ is a constant function. Nevertheless, integrating these cases separately we have checked that, in both circumstances, the solution can only be attained for $\Lambda=0$ and that, likewise, these solutions turn out to be flat spaces. Thus, summing up, the case considered in this subsection, namely $S_x'=0$ and $S_y'=0$, do not lead to any interesting solution. More precisely, all solutions in such subcase are flat.


\subsection{Subcase $S_x'\neq0$ and $S_y'=0$} \label{SubSecB}

Now, let us integrate Einstein's vacuum equation for the subcase $S_x'\neq0$ and $S_y'=0$. Since the functions $S_x$ and $S_y$ appear in the metric only through the combination $S_x + S_y$, it follows that the constant value of $S_y$ can be absorbed into $S_x$. Thus, without loss of generality, we can set
\begin{equation*}
  S_y = 0 \,.
\end{equation*}

Assuming that $P_y'\neq 0 $, it follows that $\Delta_y$ and $Q_y$ should be given by Eqs. (\ref{Dy2}) and (\ref{Qy}), respectively. With these at hand, it follows that integration of the component $R_{\sigma_1}^{\ph{\sigma_1}\sigma_1}- R_{y}^{\ph{y}y}=0$ of Einstein's equation yields
\begin{equation}\label{SSy}
\Omega_y = c_2 + c_3 P_y,
\end{equation}
with $c_2$ and $c_3$ being integration constants. Also, using the equation $R_{x}^{\ph{x}x}=\Lambda$, we obtain
\begin{equation}\label{D1}
\Delta_x=\frac{ c_4\, S_x^2 -  4 \Lambda \, S_x^3  }{ 3(S_x')^2 },
\end{equation}
where $c_4$ is another integration constant. Finally, imposing $R_{\sigma_1}^{\ph{\sigma_1}\sigma_1}= \Lambda$ we arrive at the following constraint on the integration parameters:
\begin{equation}\label{c2c4}
c_4 = -3 c_1(c_2+a_1 c_3)(c_2+a_2 c_3)\,.
\end{equation}
Then, once assumed that $c_4$ is given by Eq. (\ref{c2c4}), it follows that Einstein's vacuum equation $R_{a}^{\ph{a}b}=\Lambda\delta_{a}^{\ph{a}b}$ are fully obeyed. Nevertheless, one can check that this final solution has vanishing Weyl tensor, so that the Riemann tensor obeys
\begin{equation}
R_{abcd}=\frac{1}{3}\Lambda \left(g_{ac}g_{bd}-g_{ad}g_{bc}\right).
\end{equation}
Thus, the solution that we have found have constant curvature, i.e. they are de Sitter and anti de Sitter spacetimes when Lorentzian signature is assumed and $\Lambda\neq 0$, while it is the flat space for vanishing cosmological constant.

A possibility that has not been considered yet for the present subcase ($S_x'\neq0$ and $S_y'=0$) is $P_y'=0$, in which case Eqs. (\ref{Dy2}) and (\ref{Qy}) are not valid. However, integrating Einstein's equation for $S_y=0$ and $P_y'=0$ we also eventually find that the solution is a maximally symmetric space. Thus, all solutions of the subcase $S_x'\neq0$ and $S_y'=0$ turn out to be the ``non-interesting'' spaces of constant curvature.


\subsection{Subcase $S_x'=0$ and $S_y'\neq0$} \label{SubSecC}

Finally, let us consider the subcase $S_x'=0$ and $S_y'\neq 0$, in which case we can, without loss of generality, absorb the constant value of $S_x$ into $S_y$ and set
\begin{equation}\label{Sx3}
  S_x = 0 \,.
\end{equation}
Moreover, we can easily redefine the coordinate $x$ ($dx\rightarrow d\tilde{x} = dx/\sqrt{\Delta_x}$) in order to eliminate the function $\Delta_x$. Doing so, and dropping the tilde over the new coordinate, we find that this is equivalent to setting
\begin{equation}\label{Dx3}
 \Delta_x = 1 \,.
\end{equation}
In particular, note that due to Eqs. (\ref{Sx3}) and (\ref{Dx3}) the metric is independent of the coordinate $x$. Thus, besides the Killing vector fields $\partial_{\sigma_1}$ and $\partial_{\sigma_2}$, $\partial_{x}$ does also generate a symmetry. These three independent Killing vector fields commute with each other and, therefore, yields an abelian three-dimensional algebra. According to Bianchi's classification of three-dimensional Lie Algebras, this isometry algebra is of Bianchi type I  \cite{LBianchi}. Moreover, it is worth noting that the Killing tensor (\ref{KillingT1}) is trivial in this subcase. Indeed, with the choices (\ref{Sx3}) and (\ref{Dx3}) we get $\bl{K}=-\partial_x^2$, so that the first integral associated to $\bl{K}$ for the geodesic motion  is just the square of the one associated to the Killing vector $\partial_{x}$ \cite{Santillan}.

Postponing the analysis of the special case in which $P_y$ is constant, we can assume expressions (\ref{Dy}) and (\ref{Qy}) to hold. Doing so, and using \eqref{Sx3} and \eqref{Dx3}, it follows from the integration of $R_{\sigma_1}^{\ph{\sigma_1}\sigma_1}- R_{x}^{\ph{x}x}=0$ that $\Omega_y$ must be given by
\begin{equation}\label{Omegay3}
\Omega_y=c_2(P_y-a_1)^{d}(P_y-a_2)^{1-d},
\end{equation}
where $c_2$ and $d$ are arbitrary integration constants. Then, assuming \eqref{Omegay3} to hold, it follows from the integration of  $R_{\sigma_1}^{\ph{\sigma_1}\sigma_1}- R_{y}^{\ph{y}y}=0$ that
\begin{equation}\label{S2}
S_y=\left[b_1\left(\frac{P_y-a_1}{P_y-a_2}\right)^{d_+} + b_2\left(\frac{P_y-a_1}{P_y-a_2}\right)^{d_-}\right]^{-2/3}.
\end{equation}
In the above expression, while $b_1$ and $b_2$ are, for the time being, arbitrary integration constants, $d_{\pm}$ are not arbitrary,
rather they are given in terms of $d$ by:
\begin{equation}\label{epm}
  d_{\pm} =\frac{1}{2}\left[1 - 2\,d \pm \sqrt{d(d-1)+1}\right] \,.
\end{equation}
Finally, integrating $R_x^{\ph{x}x}=\Lambda$, we find that $b_1$ and $b_2$ must be constrained by the following relation:
\begin{equation}\label{b1b2}
b_1 b_2 = \frac{3\Lambda}{c_1 c_2^2(a_1-a_2)^2[d(d-1)+1]} \,.
\end{equation}
In order for the latter expression to be meaningful we need to have $a_1\neq a_2$. The special case $a_1=a_2$ will be considered latter. With the above prescriptions, namely Eqs. (\ref{Dy}), (\ref{Qy}), and (\ref{Sx3})-(\ref{b1b2}), we have that Einstein's vacuum equation is fully obeyed. Notice that in this solution the function $P_y$, apart from being non-constant,  has not been constrained. This freedom on the choice of $P_y$ is expected from the fact that in the metric \eqref{metric2} we could, for instance, have set $\Delta_y=1$ by means of a redefinition of the coordinate $y$. Thus, we have started with more degrees of freedom than necessary. The important point is that different choices of $P_y$ can be understood as different choices of the coordinate $y$ and, therefore, represent the same physical space.


\subsubsection{Turning the metric into a diagonal form}\label{SubSecC1}


Now, let us try to identify the solution just found.
Integrating the Killing equation, we can check that this solution admits no other independent generators of symmetries besides the commuting Killing vector fields $\partial_{\sigma_1}$,  $\partial_{\sigma_2}$, and $\partial_{x}$. Thus, this solution is, indeed, a Bianchi Type I space.

A well-known class of spacetimes that are Bianchi type I are the so-called \textit{Bianchi type I cosmological spacetimes}, which have the diagonal form
\begin{equation}\label{bianchiI}
ds^2=-d\tau^2+(A^1_{\tau})^2dz_1^2+ (A^2_{\tau})^2   dz_2^2+ (A^3_{\tau})^2   dz_3^2 \,,
\end{equation}
where $A^1_{\tau}$, $A^2_{\tau}$ and $A^3_{\tau}$ are arbitrary functions of $\tau$. These spacetimes are used by cosmologists to incorporate anisotropy at the space-like hyper-surfaces $\tau=constant$, providing a generalization of the FRW cosmological model \cite{Jacobs}. The Killing vectors $\partial_{z_1}$, $\partial_{z_2}$ and $\partial_{z_3}$ generate a three-dimensional abelian isometry group, so that the isometry algebra is of Bianchi type I. This isometry group acts transitively on the three-dimensional hyper-surfaces given by $\tau = constant$. The diagonal form of this line element indicates that the coordinate vector fields are orthogonal to family hyper-surfaces. In particular, the Killing vectors $\partial_{z_1}$, $\partial_{z_2}$, and $\partial_{z_3}$  are hyper-surface orthogonal. For instance, $\partial_{z_1}$ is orthogonal to the hyper-surfaces $z_1=constant$.

Coming back to our Bianchi type I solution found in the present subsection, one can see that while $\partial_x$ is a hyper-surface orthogonal Killing vector, the existence of the term $d\sigma_1 d\sigma_2$ in the line element (\ref{metric2}) indicates that the Killing vector fields $\partial_{\sigma_1}$ and $\partial_{\sigma_2}$ are not orthogonal to families of hyper-surfaces. Indeed, we can check that
\begin{equation*}
(\partial_{\sigma_1})_{[a}\nabla_b(\partial_{\sigma_1})_{c]}\neq0 \quad \text{and} \quad
(\partial_{\sigma_2})_{[a}\nabla_b(\partial_{\sigma_2})_{c]}\neq0 \,.
\end{equation*}
Thus, let us try to find two independent Killing vector fields that are orthogonal to families of hyper-surfaces to replace $\partial_{\sigma_1}$ and $\partial_{\sigma_2}$. Defining the Killing vector field
\begin{equation*}
  \bl{k} = \alpha\,\partial_{\sigma_1} + \partial_{\sigma_2}
\end{equation*}
and imposing the condition $k_{[a}\nabla_bk_{b]}=0$, one can find that as long as $Q_y=\sqrt{(P-a_1)(P-a_2)}$, irrespective of form of the other functions appearing in the line element (\ref{metric2}), we end up with two possible values for the constant parameter $\alpha$:  either $\alpha=a_1$ or $\alpha=a_2$. Thus, whenever $a_1\neq a_2$ we can exchange the independent Killing vector fields $\partial_{\sigma_1}$ and $\partial_{\sigma_2}$ by
\begin{equation}\label{k1k2}
\bl{k_1} = a_1\partial_{\sigma_1}+\partial_{\sigma_2} \quad \text{and} \quad \bl{k_2} = a_2\partial_{\sigma_1}+\partial_{\sigma_1}\,,
\end{equation}
which are also independent if $a_1\neq a_2$. The important point is that $\bl{k_1}$ and $\bl{k_2}$ are hyper-surface orthogonal, differently from $\partial_{\sigma_1}$ and $\partial_{\sigma_2}$.
 Since $\bl{k_1}$ and $\bl{k_2}$ commute with each other, we can associate to them coordinates $\phi_1$ and $\phi_2$ such that $\bl{k_1} = \partial_{\phi_1}$ and $\bl{k_2} = \partial_{\phi_2}$. Indeed, $\phi_1$ and $\phi_2$ are defined by
\begin{equation}
\sigma_1 = a_1 \phi_1 + a_2 \phi_2 \quad \text{and} \quad \sigma_2= \phi_1 +\phi_2 \,.
\end{equation}
In terms of these coordinates, the line element (\ref{metric2}) takes the form below:
\begin{align}
ds^2=& \frac{S_y}{\Delta_y}dy^2   -\frac{(a_2-a_1)(P_y-a_1)S_y}{\Omega_y}d\phi_1^2 \nonumber \\
&+\frac{(a_2-a_1)(P_y-a_2)S_y}{\Omega_y}d\phi_2^2 + S_ydx^2. \label{metric4}
\end{align}
This diagonal line element can be easily put in the general form (\ref{bianchiI}) by redefining the coordinate $y$.

The fact that the investigated solution could be diagonalized using three cyclic coordinates, $\phi_1$, $\phi_2$ and $x$, could be anticipated from the fact that if we take a general Killing vector field, $\bl{\eta} = \lambda_1   \partial_{\sigma_1} + \lambda_2   \partial_{\sigma_2} + \lambda_3   \partial_{x} $ and compute its squared norm, we will conclude that if $a_1\neq a_2$ then $\eta^a\eta_a = 0$ only if $\lambda_1=\lambda_2=\lambda_3 = 0$. Thus, the hyper-surfaces $y=constant$, spanned by the  Killing vector fields, have metrics that are either positive-definite or negative-definite. In this circumstance, there is a result on the literature stating that the metric can be diagonalized.   Indeed, in \cite{Jacobs} it is shown that it is always possible to diagonalize a metric of the form $ds^2=-dt^2+\gamma_{ij}dx^i dx^j$, where $\gamma_{ij}$ is a positive/negative-definite three-dimensional metric, whenever the Einstein's vacuum equation with cosmological constant is imposed. Nevertheless, for the case in which $a_1=a_2$ we can have a non-zero light-like Killing vector, so that the diagonalization cannot be attained using cyclic coordinates.

Remember that the non-constant function $P_y$ has not been constrained, which was a consequence of the freedom in the choice of the coordinate $y$, as argued above. Thus, without any loss of generality, we can set
\begin{equation}\label{Py}
  P_y = \frac{a_2 F_y -a_1}{F_y -1} \,,
\end{equation}
with the function $F_y$ being defined by
\begin{equation*}
F_y=\left[\sqrt{\frac{b_2}{b_1}}\tan\left(\frac{\sqrt{3\Lambda}y}{2}\right)\right]^{2/\sqrt{d(d-1)+1}}.
\end{equation*}
This choice in the coordinate $y$ was made so that the component $g_{yy}$ of the metric became equal to the unit. Then, assuming (\ref{Py}) to hold and replacing the cyclic coordinates $\phi_1$, $\phi_2$ and $x$ by their rescaled versions defined by
\begin{align*}
  x_1 &=  \sqrt{ \frac{(3\Lambda)^{p_1}(a_1-a_2)    b_2^{p_1-2/3}}{2^{2p_1} \,c_2\, b_1^{p_1}}}   \,\, \phi_1 \,,\\
  x_2 &=   \sqrt{\frac{(3\Lambda)^{p_2}     (a_2-a_1)    b_2^{p_2-2/3}}{2^{2p_2} \,c_2\, b_1^{p_2}}}  \,\, \phi_2 \,, \\
  x_3 &=   \sqrt{\frac{(3\Lambda)^{p_3}    b_2^{p_3-2/3} }{ 2^{2p_3} \  b_1^{p_3}}}   \,\, x \,,
\end{align*}
with the constant parameters $p_i$ given by
\begin{align*}
p_1&=\frac{2-d}{3\sqrt{d(d-1)+1}} + \frac{1}{3}\,,\\
p_2 &=-\,\frac{d+1}{3\sqrt{d(d-1)+1}} + \frac{1}{3} ,\\
p_3&=\frac{2d-1}{3\sqrt{d(d-1)+1}} + \frac{1}{3} ,
\end{align*}
it follows that the line element (\ref{metric4}) becomes
\begin{equation}\label{KasnerM}
ds^2=dy^2+L_y^{2/3}\Bigg[ \sum_{i=1}^{3}e^{2(p_i-\frac{1}{3})N_y}(dx_i)^2\Bigg]\,,
\end{equation}
where
 \begin{equation*}
L_y =\frac{\sin(\sqrt{3\Lambda}y)}{\sqrt{3\Lambda}} \, ,\;\;
N_y =\textrm{Log}\left[\frac{2 \,\tan\left( \sqrt{3\Lambda}y/2\right)}{\sqrt{3\Lambda}} \right].
\end{equation*}
Note that the parameters $p_i$ obey the following constraint.
\begin{equation}
\sum_{i=1}^{3} p_i= 1\,, \; \textrm{ and } \; \sum_{i=1}^{3} p_i^2 = 1.
\end{equation}
The solution (\ref{KasnerM}) is a generalization of the Kasner metric for the case in which the cosmological constant is different from zero. This particular solution is already available in the literature, see chapter 13 of Ref. \cite{Stephani}. In the limit $\Lambda\rightarrow 0$ the solution becomes
\begin{equation}
ds^2=dy^2+y^{2p_1}dx_1^2+ y^{2p_2}dx_2^2+y^{2p_3}dx_3^2,
\end{equation}
which is Kasner Metric \cite{Stephani, EKasner}. Such a solution is used in cosmology to model an anisotropic vacuum universe \cite{Jacobs}.

In order to obtain the latter solution we have avoided two special cases, namely we have assumed that $P_y$ is non-constant, so that (\ref{Dy}) hold, and have assumed $a_1\neq a_2$. Thus, for completeness, we should also tackle these cases. First, considering  $P_y$ constant and following steps analogous to the ones adopted above, one can check that solutions can be attained but all these solutions are either equivalent to (\ref{KasnerM}) or one of its subcases. So, the special case in which $P_y$ is constant does not lead to new solutions. Differently, the special case $a_1=a_2$ will yield a new solution that is not available in the literature. In the case $a_1=a_2$,  Eqs. (\ref{S2}) and (\ref{b1b2}) are not valid so that the calculations should be done separately, which we shall do in the next subsection. Note that in this special case the Killing vectors (\ref{k1k2}) are not independent from each other, so that the diagonal form above cannot be attained, as hinted by the fact that the coordinates $\phi_1$ and $\phi_2$ are proportional to each other when $a_1=a_2$.


\subsubsection{The special case $a_1=a_2$} \label{SubSecC2}

The special case $a_1=a_2$ will be considered in the present section. It turns out that this will be the most interesting case, since, as far as the authors know, the obtained solution has not been described in the literature yet.

In the sequel, we will assume
\begin{equation}\label{SxDxDy}
   S_x=0 \;,\;\; \Delta_x=1 \;,\; \textrm{and } \; \Delta_y=\frac{c_1Q_y^2\Omega_y^2}{S_y^2(P_y')^2}\,,
\end{equation}
as assumed for the general case, whereas the function $Q_y$ reduces to
\begin{equation}\label{delta2Q}
 Q_y = P_y-a_1 \,,
\end{equation}
since now $a_1=a_2$. Then, from the integration of the equation $R_{\sigma_1}^{\ph{\sigma_1}\sigma_1}- R_{x}^{\ph{x}x}=0$, we obtain
\begin{equation}\label{Sk}
\Omega_y=c_2\,Q_y\,e^{-\tilde{d}/Q_y},
\end{equation}
where $c_2$ and $\tilde{d}$ are arbitrary integration constants.
Using this result for integrating the equation $R_{\sigma_1}^{\ph{\sigma_1}\sigma_1}- R_{y}^{\ph{y}y}=0$, we find that
\begin{equation}\label{S2k}
S_y=\left[ b_1 e^{3\tilde{d}/(2Q_y)} + b_2\, e^{\tilde{d}/(2Q_y)} \right]^{-2/3},
\end{equation}
with $b_1$ and $b_2$ being arbitrary integration constants.
Finally, solving  $R_{\sigma_1}^{\ph{\sigma_1}\sigma_1}=\Lambda$, we conclude that the constants $b_1$ and $b_2$ must be related to $\Lambda$ as follows:
\begin{equation}\label{b1b2k}
b_1 b_2=\frac{3\Lambda}{ c_1 \,c_2^2 \, \tilde{d}^2}.
\end{equation}
This concludes the integration, as it can be checked that the remaining components of Einstein's equations are obeyed. Thus, we have completely integrated Einstein's equations for the particular case in which $a_1=a_2$, the general solution being given by the line element (\ref{metric2}) with its functions given by (\ref{SxDxDy})-(\ref{b1b2k}). An interesting fact is that this solution for the case $a_1=a_2$ can be obtained from the case $a_1\neq a_2$ by defining
\begin{equation*}
 d = \frac{\tilde{d}}{a_1 - a_2}
\end{equation*}
and then taking the singular limit $a_2\rightarrow a_1$ in the expressions (\ref{Omegay3}), (\ref{S2}), and (\ref{b1b2}).

Now, let us try to put the solution just found in a neater form. First, let us make use of the degree of freedom on the choice of $P_y$ to set
\begin{equation*}
  P_y = y \,.
\end{equation*}
As explained before, this amounts to no loss of generality. Then, we shall perform the coordinate transformation $(\sigma_1, \sigma_2, x, y)\rightarrow (t,\phi,\theta,r)$, where the new coordinates are defined by
\begin{align*}
  \sigma_1 &= -    \frac{\sqrt{c_2} \, b_1^{1/3}}{2\sqrt{\tilde{d}} } \left[ (\tilde{d}+ a_1\,\tilde{c}) \,t - a_1 \phi\right] \\
\sigma_2 &=    \frac{\sqrt{c_2} \, b_1^{1/3}}{2 a_1 \sqrt{\tilde{d}} } \left[ (\tilde{d}- a_1\,\tilde{c}) \,t + a_1 \phi\right]\\
  x&=   b_1^{1/3} \, e^{(a_1\tilde{c}-\tilde{d})/(2a_1)} \,\, \theta \,,\\
y &= \frac{a_1^2\,(r + \tilde{c} )}{a_1 \, r + a_1\tilde{c}-\tilde{d}} \, ,
\end{align*}
with the constant $\tilde{c}$ standing for
\begin{equation*}
   \tilde{c} = \frac{\tilde{d}}{a_1}-\log(c_1\,c_2^2\,b_1^2\, \tilde{d}^2 /3)\,.
\end{equation*}
In terms of these new coordinates the line element is given by
\begin{equation}\label{metric23}
ds^2=\frac{e^{-r} dr^2}{3(1+\Lambda e^{-r})^2}+\frac{e^{-r}d\theta^2-dt(r\, dt+d\phi)}{(1+  \Lambda e^{-r} )^{2/3}}\,.
\end{equation}
Notice that we were able to get rid of all of the integration constants, so that this solution depends just on the cosmological constant, which is an external parameter. In these coordinates the metric is Lorentzian, although the signature could be easily changed by means of Wick rotations.


\section{Analyzing The New Solution }\label{Sec.NewSOL}

In this section we shall analyze the geometrical properties of the line element (\ref{metric23}) aiming the identification of the spacetime. As we will argue in the sequel, such analysis hints that the metric given in Eq. (\ref{metric23}) might be a new exact solution for Einstein's vacuum equation. In order to arrive at this conclusion, we have tried to characterize this line element as much as possible and then looked for known solutions with the same geometrical features. The bottom line is that as far as the authors were able to investigate, the solution (\ref{metric23}) has not been defined in the literature yet.

First, let us point out that the special case of vanishing cosmological constant of the solution (\ref{metric23}) is already described in the literature. Indeed, when $\Lambda=0$ it follows that $\partial_\phi$ is a covariantly constant null vector field, so that the line element represents a $pp$-wave spacetime \cite{Stephani,EhlersKundt}. The $pp$-wave spacetimes are Petrov type $N$ and all their curvature scalars vanish identically (VSI spacetimes), for more in this class of spaces see Ref. \cite{VPravdaEtAl}.

Thus, it remains to analyze the general case $\Lambda\neq0$.  A good starting point is to investigate the isometry group of the solution (\ref{metric23}). A complete integration of the Killing equation yields that the isometry group is three-dimensional and abelian, with the trivial Killing vectors $\partial_t$, $\partial_\theta$,  and $\partial_\phi$ being a basis for the isometry Lie algebra. So, the isometry algebra is of Bianchi type I. Forming a general linear combination of these Killing vectors, we can see that the only ones that are orthogonal to families of hyper-surfaces are $\partial_\theta$ and $\partial_\phi$.  Moreover, note that the Killing vector field $\partial_\phi$ is null. In particular, the existence of a null Killing vector implies that the line element cannot be put in a diagonal form using cyclic coordinates, differently from the previous case $a_1\neq a_2$, see the discussion on the paragraph below Eq. (\ref{metric4}).

Besides studying the isometry group, another geometric way to characterize the solution \eqref{metric23} is analysing its Petrov type. In order to do so, we need to use a so-called null tetrad frame $\{\bl{\ell},\bl{n},\bl{m},\bl{\bar{m}}\}$, in which the vector fields $\bl{\ell}$ and $\bl{n}$ are real, while $\bl{m}$ and $\bl{\bar{m}}$ are complex and conjugated to each other. The only non-vanishing inner products in such a frame are
$\ell^a n_{a} = -1$ and $m^a \bar{m}_{a} = 1$. Using one of the null tetrad frames below, i.e. choosing either the $+$ frame or the $-$ frame,
\begin{widetext}
\begin{align*}
 \bl{\ell} &=  \partial_\phi \,,  \\
 \bl{n}_{\pm} &= \pm\frac{e^r\sqrt{2}}{\sqrt{\Lambda}} (1+  \Lambda e^{-r})^{2/3}(3+ \Lambda e^{-r})^{1/2} \, \partial_\theta \
 + 2 (1+  \Lambda e^{-r})^{2/3} \partial_t
+ \frac{1}{  \Lambda } (1+  \Lambda e^{-r})^{2/3}[3 e^r + \Lambda(1-2r) ] \, \partial_\phi \,, \\
 \bl{m}_{\pm} &= \frac{\sqrt{3}e^{r/2}}{\sqrt{2\,}} (1+  \Lambda e^{-r}) \, \partial_r \
 + i\,\frac{e^{r/2}}{\sqrt{2}} (1+  \Lambda e^{-r})^{1/3} \partial_\theta
\pm i\, \frac{e^{r/2}}{\sqrt{\Lambda}} (3+  \Lambda e^{-r})^{1/2} (1+  \Lambda e^{-r})^{1/3} \, \partial_\phi  \,,  \\
\bl{\bar{m}}_{\pm} &= \frac{\sqrt{3}e^{r/2}}{\sqrt{2\,}} (1+  \Lambda e^{-r}) \, \partial_r \
 - i\,\frac{e^{r/2}}{\sqrt{2}} (1+  \Lambda e^{-r})^{1/3} \partial_\theta
\mp i\, \frac{e^{r/2}}{\sqrt{\Lambda}} (3+  \Lambda e^{-r})^{1/2} (1+  \Lambda e^{-r})^{1/3} \, \partial_\phi \,,
\end{align*}
\end{widetext}
it follows that the only Weyl scalars different from zero are, respectively,
\begin{align*}
  \Psi_2 &= \frac{\Lambda}{6} (1+ \Lambda e^{-r}) \, ,\, \textrm{ and} \\
   \Psi_3 &=\mp \,i  \sqrt{\frac{\Lambda e^r}{4 }} (1+ \Lambda e^{-r})^{4/3} (3+ \Lambda e^{-r})^{1/2}\,.
\end{align*}
The fact that $\Psi_0$, $\Psi_1$, and $\Psi_4$ all vanish in these frames means that $\bl{\ell}=\partial_\phi$ is a repeated principal null direction of the Weyl tensor, while $\bl{n}_{\pm}$ are non-degenerated principal null directions. Moreover, this implies that the Weyl tensor is of Petrov type II.  For some review on the Petrov classification, see Ref. \cite{Bat-Book-art2}.

Another important geometric characterization of this spacetime is that the null vector field  $\partial_\phi$ is geodesic, shear-free, twist-free, and expansion-free. This means that the above solution is contained in the Kundt class of spacetimes. For a recent review on this class of spacetimes see   \cite{ColeyPapadopoulos}.

All the above features of the solution (\ref{metric23}) have been extensively used in order to try to find it in the literature. In particular, a thorough search has been performed by the authors on the books \cite{Stephani,GrifPodol-Book}. In fact, the closest that we could get from finding such a solution in the literature was in chapter 31 of Stephani et. al. book \cite{Stephani}, where they exhibit the Kundt's class of spacetimes. In particular, for solutions of Petrov type II with non-zero cosmological constant, the authors of \cite{Stephani} refer to two papers, \cite{Garcia} and \cite{Khlebnikov}, where special solutions in such a class of spacetimes are found. However, our solution (\ref{metric23}) could not be found there, inasmuch as their solutions contain strictly nonzero electromagnetic fields. In light of this, it seems to the authors of the present paper that the spacetime described by the metric (\ref{metric23}) has not been presented in the literature so far, being a new solution of Einstein's field equations with cosmological constant. Actually, the analysis of the existing literature revealed that there are few known exact vacuum solutions of Petrov type II. In contrast, solutions of Petrov type D are much more abundant. For instance, W. Kinnersley has been able to fully integrate Einstein's vacuum equation with vanishing cosmological constant for the entire class of type D spacetimes \cite{typeD}, yielding a plethora of solutions, a particular example being Kerr metric.

Concerning the regularity of the line element (\ref{metric23}), it seems that it is regular at all range of the coordinate $r$ except for the point $r=-\infty$ and when the denominator $(1+\Lambda e^{-r})$ vanishes. Computing some curvature scalars we have found the following pattern:
\begin{align}
  & R^{a_1b_1}_{\ph{a_1b_1}a_2b_2}  R^{a_2b_2}_{\ph{a_1b_1}a_3b_3} \cdots R^{a_nb_n}_{\ph{a_nb_n}a_1b_1} =   \nonumber\\
 & \quad  4 \sum_{j=0}^{n}\,\binom{n}{j}\, \frac{e^{-jr} }{3^j} \Lambda^{n+j} + \frac{2}{3^n}  (-2 \Lambda^2 e^{-r})^n\,, \label{Scalars2}
\end{align}
where $R_{abcd}$ stands for the Riemann tensor.
Note that all these scalars are finite for $r\neq - \infty$. On the other hand, in the limit
$r\rightarrow- \infty$ these scalars diverge exponentially like $e^{n|r|}$. Thus, the point $r=-\infty$ is a singularity of the spacetime, while other points are regular. Likewise, computing the curvature scalar
\begin{equation}\label{DR2}
  \nabla^{a}R^{bcde}\nabla_{a}R_{bcde} = \frac{20}{3}  \Lambda^4 (1+\Lambda e^{-r})^2 e^{-r}\,,
\end{equation}
we check that there is a divergence just at $r=-\infty$.

Note that when the cosmological constant is negative the denominator $(1+\Lambda e^{-r})$  can vanish, which could indicate the existence of a real singularity at $r= \log(-\Lambda)$, inasmuch as the line element (\ref{metric23}) blows up. However, the fact that the curvature scalars (\ref{Scalars2}) and (\ref{DR2}) are perfectly regular at $r= \log(-\Lambda)$ reveals that this is not the case. In other words, the divergence of the metric components at $r= \log(-\Lambda)$, when $\Lambda < 0$, is just a coordinate singularity.

The asymptotic limit $r\rightarrow \infty$ has a particularly simple structure concerning the curvature scalars. While Eq. (\ref{DR2}) reveals that the square of the derivative of the curvature tensor goes to zero in this limit, the powers of the Riemann tensor given in Eq. (\ref{Scalars2}) goes to $4\Lambda^n$ when $r\rightarrow \infty$. Such a simple structure reminds of spaces of constant curvature like (anti-)de Sitter, $(a)dS_4$, which is a four-dimensional Lorentzian space of constant curvature, and (anti-)Nariai, $(a)N_4$, which is a solution of Einstein's equation that is the direct product of two spaces of constant curvature.  However, although these two spacetimes have covariantly constant Riemann tensors, so that $\nabla^{a}R^{bcde}\nabla_{a}R_{bcde}=0$, in agreement with the behaviour of Eq. (\ref{DR2}) in the limit $r\rightarrow \infty$, the powers of the Riemann tensor differ from the ones of our spacetime. Instead of $4\Lambda^n$, which is obtained from Eq. (\ref{Scalars2}) in the limit $r\rightarrow \infty$, for these solutions we have
\begin{equation*}
   R^{a_1b_1}_{\ph{a_1b_1}a_2b_2} \cdots R^{a_nb_n}_{\ph{a_nb_n}a_1b_1} = \left\{
                                                                            \begin{array}{ll}
                                                                             (a)dS_4:\;\; 6 (2\Lambda/3)^n \,, \\
                                                                              \quad \\
                                                                              (a)N_4:\;\; 2 (2\Lambda)^n\,.
                                                                            \end{array}
                                                                          \right.
\end{equation*}
Thus, we can state that the new solution is neither asymptotically $(a)dS_4$ nor asymptotically $(a)N_4$.

In order to investigate the asymptotic limit of our solution, we shall focus on the block related to $dt$ and $d\phi$ in the line element (\ref{metric23}), namely let us consider
\begin{equation*}
  ds^2_{t\phi} \equiv - dt(\,r \,dt + d\phi) \,.
\end{equation*}
Then, performing the coordinate transformation $(t,\phi) \rightarrow (\tilde{t},\tilde{\phi})$, where
\begin{equation*}
  \tilde{t} = r\,t \,, \;\; \textrm{and} \;\; \tilde{\phi} = r^{-1}\,\phi \,,
\end{equation*}
it follows that $ds^2_{t\phi}$ becomes:
\begin{equation*}
  ds^2_{t\phi} = -d\tilde{t} \, d\tilde{\phi} - \frac{1}{r}\left[ d\tilde{t}^2 + \tilde{\phi}\, d\tilde{t} dr  - \tilde{t}\, d\tilde{\phi}dr \right] + O\left( r^{-2} \right),
\end{equation*}
where $O\left(r^{-2}\right)$ denotes terms that fall off as $r^{-2}$, or faster, when $r\rightarrow\infty$.
Thus, in terms of the coordinates $(\tilde{t},\tilde{\phi})$, the asymptotic limit of the block $ds^2_{t\phi}$ becomes
\begin{equation*}
  ds^2_{t\phi}|_{r\rightarrow\infty} \simeq -d\tilde{t} \, d\tilde{\phi} \,.
\end{equation*}
Hence, we can say that in the asymptotic limit the solution (\ref{metric23}) converges to
\begin{equation}\label{metric23-Limit_1}
 ds^2|_{r\rightarrow\infty} \simeq \frac{ e^{-r} dr^2}{3(1+ \Lambda e^{-r})^2}+
\frac{e^{-r}d\theta^2-d\tilde{t}\,d\tilde{\phi} }{(1+  \Lambda e^{-r} )^{2/3}}\,.
\end{equation}
This limit spacetime turn out to be a particular member of the generalized Kasner class of solutions, as demonstrated in App. \ref{AppendixA}. More precisely, the solution (\ref{metric23-Limit_1}) corresponds to the choice $(p_1,p_2,p_3)=(2/3,2/3,-1/3)$ of the generalized Kasner metric (\ref{KasnerM}). As shown in App. \ref{AppendixA}, this limit spacetime is of Petrov type D and possess a four-dimensional isometry algebra. Curiously, one can check that the curvature scalars of the line element (\ref{metric23-Limit_1}) are exactly the same as the ones of the solution (\ref{metric23}), namely Eqs. (\ref{Scalars2}) and (\ref{DR2}) are also valid for the solution (\ref{metric23-Limit_1}). This, however, do not imply that these two spacetimes are the same. Indeed, it is well-known that two geometries can have the same curvature scalars and still be different from each other \cite{Coley:2009eb,Olver}. A famous example is given by $pp$-wave spacetimes, which, in spite of having all curvature scalars equal to zero, are not flat. Thus, here we have obtained another example of two different spacetimes with the same curvature scalars.

\section{Conclusions and Perspectives}\label{Sec.Conc}

In this paper we have completely integrated Einstein's vacuum equation with a cosmological constant for a subclass of the most general four-dimensional metric containing two commuting Killing vector fields and a non-trivial Killing tensor of rank two. As we have seen, most of the solutions then found have already been described in the literature. Among them, we have obtained flat space, spaces of constant curvature, and a generalization of the Kasner metric to the case of non-zero cosmological constant. Nevertheless, we have also obtained a solution that, as far as we know, have never been described in the literature before, see Eq. (\ref{metric23}). In order to arrive at this conclusion some  features of this solution were investigated, such as its isometry group, its Petrov type, and the optical scalars related to the null Killing vector field of this solution. More precisely,  we have obtained that the isometry algebra of this solution is three-dimensional and abelian, which means that it is Bianchi type I, its Weyl tensor is of Petrov type II, and the solution is contained in the Kundt class of spacetimes. Then, we searched in the literature pre-existing vacuum solutions having the same features, but no match occurred. Finally, we have proved that in the asymptotic limit $r\rightarrow\infty$ this solution approaches a member of the class of generalized Kasner spacetimes which have the same curvature scalars.

We hope that this new solution, along with the characterization given in this paper could give rise to applications within the framework of gravitation, cosmology and beyond. The analysis of the physical properties of the solution (\ref{metric23}) can give a hint on the range of its applicability. Therefore, in a future work we intend to investigate the physics of such exact solution .

\begin{acknowledgments}
C. B. would like to thank Conselho Nacional de Desenvolvimento Cient\'{\i}fico e Tecnol\'ogico (CNPq) for the partial financial support through the research productivity fellowship. Likewise,  C. B. thanks Universidade Federal de Pernambuco for the funding through Qualis A project.  G. L. A. thanks CNPq for the financial support.
\end{acknowledgments}

\appendix

\section{Another Solution Through a Singular Limit}\label{AppendixA}

In this appendix we shall investigate some geometric properties of the spacetime (\ref{metric23-Limit_1}), which is the asymptotic limit of the new solution (\ref{metric23}) when $r\rightarrow\infty$. In particular, we  will prove that its Weyl tensor is type D according to the Petrov classification and that its isometry group is four-dimensional.

Let us start proving that the spacetime (\ref{metric23-Limit_1}) can be obtained from the new solution (\ref{metric23}) by means of a singular coordinate transformation.
Replacing the coordinates $t$ and $\phi$ in the line element (\ref{metric23})  by $\tilde{t}$ and $\tilde{\phi}$ defined as
\begin{equation}\label{t phi LimitMetric}
  \tilde{t} = \lambda^{-1/2}\,t \;\, \textrm{ and } \,\; \tilde{\phi} = \lambda^{1/2}\,\phi \,,
\end{equation}
where $\lambda$ is a positive constant parameter, we are led to
\begin{equation*}
ds^2=\frac{ e^{-r} dr^2}{3(1+\Lambda e^{-r})^2}+
\frac{e^{-r}d\theta^2-d\tilde{t}(\,\lambda \,r\, d\tilde{t}+d\tilde{\phi})}{(1+  \Lambda e^{-r} )^{2/3}}\,.
\end{equation*}
Note that although the limit $\lambda\rightarrow0$ is forbidden at the level of the coordinates, since $\tilde{t}$ and $\tilde{\phi}$ become ill-defined, the line element obtained in this limit is perfectly regular and is given by
\begin{equation}\label{metric23-Limit}
ds^2=\frac{ e^{-r} dr^2}{3(1+ \Lambda e^{-r})^2}+
\frac{e^{-r}d\theta^2-d\tilde{t}\,d\tilde{\phi} }{(1+  \Lambda e^{-r} )^{2/3}}\,.
\end{equation}
Despite the line element (\ref{metric23-Limit}) being obtained from our solution through a coordinate transformation, the metric  (\ref{metric23-Limit}) can represent a completely different spacetime, since the coordinate transformation
\eqref{t phi LimitMetric} is singular at $\lambda=0$. For instance, another example of singular coordinate transformations that yield a different space is provided by Nariai spacetime, which can be obtained from the degenerated Schwarzschild-dS solution\footnote{By degenerated we mean that the event horizon and the cosmological horizon coincide.} by means of a singular coordinate transformation \cite{BatistaNariai}.

Now, let us investigate some properties of the solution (\ref{metric23-Limit}). Note that besides being invariant under translations in the coordinates $\theta$, $\tilde{t}$ and $\tilde{\phi}$, the line element  (\ref{metric23-Limit}) is also invariant under the boost transformation
$$ r\rightarrow r \,,\;\; \theta \rightarrow \theta \,,\;\; \tilde{t}\rightarrow a\,\tilde{t} \,,\;\; \tilde{\phi}\rightarrow \frac{1}{a}\,\tilde{\phi} \,,$$
with $a$ being an arbitrary constant parameter.  This is an extra symmetry, whose generator is the Killing vector field
$$ \bl{\tilde{k}} = \tilde{t}\, \partial_{\tilde{t}} - \tilde{\phi}\, \partial_{\tilde{\phi}}  \,. $$
One can check that this is the only extra independent killing vector of the solution (\ref{metric23-Limit})  besides the obvious ones $\partial_\theta$,  $\partial_{\tilde{t}}$, and $\partial_{\tilde{\phi}}$, so that the isometry group is four-dimensional and nonabelian. In particular, this implies that, in spite of having the same curvature scalars, the solutions  (\ref{metric23}) and (\ref{metric23-Limit}) represent different spacetimes, since they have different isometry groups.



In order to obtain the Petrov classification of the solution (\ref{metric23-Limit}), let us introduce the following null tetrad
\begin{align*}
  \bl{\ell} &= \partial_{\tilde{\phi}} \,,\\
  \bl{n} &= 2(1+  \Lambda e^{-r})^{2/3}\partial_{\tilde{t}} \,,\\
  \bl{m} &= \sqrt{\frac{3 e^{r}}{2}}\,(1+  \Lambda e^{-r})\partial_{r} + i \, \sqrt{\frac{e^{r}}{2}}\,(1+  \Lambda e^{-r})^{1/3}\partial_{\theta}\,,\\
\bl{\bar{m}} &= \sqrt{\frac{3 e^{r}}{2}}\,(1+  \Lambda e^{-r})\partial_{r} - i \, \sqrt{\frac{e^{r}}{2}}\,(1+  \Lambda e^{-r})^{1/3}\partial_{\theta}\,.
\end{align*}
Then, computing the Weyl scalars with such tetrad, we find that $\Psi_0$, $\Psi_1$, $\Psi_3$, and $\Psi_4$ vanish, while, for $\Lambda\neq 0$, $\Psi_2$ is different from zero and given by
\begin{equation*}
  \Psi_2 = \frac{\Lambda}{6} (1+  \Lambda e^{-r})\,.
\end{equation*}
This means that $\bl{\ell}$ and $\bl{n}$ are both repeated principal null directions and that the solution (\ref{metric23-Limit}) is of Petrov type D if $\Lambda\neq 0$, which differs from the Petrov classification  of the line element (\ref{metric23}).

Note also that in the case $\Lambda=0$ the latter tetrad is well-defined, so that we can use it to compute the Weyl scalars. Doing so, we see that, when $\Lambda$ is zero, $\Psi_2$ vanishes along with the $\Psi_0$, $\Psi_1$, $\Psi_3$, and $\Psi_4$. Since all Weyl scalars vanish it follows that the Weyl tensor is identically zero, which along with the fact that the metric (\ref{metric23-Limit})  is Ricci-flat when $\Lambda=0$ means that the spacetime is flat. Thus, the case $\Lambda=0$ of the line element (\ref{metric23-Limit}) is just Minkowski spacetime.

Continuing the characterization of the limit solution (\ref{metric23-Limit}), we can also verify that the null Killing vector fields $\partial_{\tilde{t}}$  and $\partial_{\tilde{\phi}}$  are geodesic, shear-free, twist-free, and expansion-free, so that the line element  (\ref{metric23-Limit}) is contained in the Kundt class. Concerning the regularity of the solution and its asymptotic limit when $r\rightarrow \infty$, all the comments made for the solution (\ref{metric23}) remains valid for the limit solution (\ref{metric23-Limit}), since these spaces have exactly the same curvature scalars.

Finally, we can check that the solution (\ref{metric23-Limit}) is, actually, a member of the generalized Kasner solutions that we have obtained in Eq. (\ref{KasnerM}). Indeed, performing the coordinate transformation $(r,\theta,\tilde{t},\tilde{\phi})\rightarrow (y,x_1,x_2,x_3)$, where
\begin{align*}
r&= 2 \log\Big[\sqrt{\Lambda}\tan\Big(\frac{\sqrt{3\Lambda}y}{2}\Big)\Big]\\
\tilde{t}&= \Big(\frac{4}{3\Lambda}\Big)^{1/3}(i x_1+x_2)\\
\tilde{\phi}&= \Big(\frac{4}{3\Lambda}\Big)^{1/3}(i x_1-x_2)\\
\theta&= \sqrt{\Lambda}\Big(\frac{4}{3\Lambda}\Big)^{-1/6}x_3\\
\end{align*}
we can see that the line element (\ref{metric23-Limit}) takes the form (\ref{KasnerM}) with the choice $(p_1,p_2,p_3)=(2/3,2/3,-1/3)$. Thus, we can say that in the asymptotic limit $r\rightarrow\infty$, our new solution (\ref{metric23}) goes to a generalized Kasner spacetime.

\end{document}